# In-Circuit Impedance Measurement Setups of Inductive Coupling Approach: A Review

Zhenyu Zhao[1], Fei Fan[1], Huamin Jie[1], Zhenning Yang[1], Minghai Dong[2], Eng Kee Chua[1], Kye Yak See[1]

[1]School of Electrical and Electronic Engineering, Nanyang Technological University, Singapore
[2]School of Aeronautics and Astronautics, University of Electronic Science and Technology of China, Chengdu, China
zhao0245@e.ntu.edu.sg

*Abstract*—In-circuit impedance measurement provides useful information for many EMC applications. The inductive coupling approach is a promising in-circuit impedance measurement method due to its non-contact characteristics and simple on-site implementation. Many measurement setups of this approach were reported. However, a comprehensive survey and comparison of these setups have not been found in the literature. This paper reviews these setups in terms of their characteristics, limitations, and applications. In addition, recommendations for future research are also presented.

## I. Introduction

In-circuit impedance measurement plays an important role in many EMC applications [1]-[5]. Three in-circuit impedance measurement methods are widely used, namely the voltage-current (V-I) approach [6], the capacitive coupling approach [7] and the inductive coupling approach [8]. The measurement setups of the inductive coupling approach have no direct electrical contact with the energized system under test (SUT), thereby simplifying on-site implementation without posing electrical safety hazards.

To measure the in-circuit impedance of one SUT, the measurement setups of the inductive coupling approach can be mainly divided into the frequency-domain two-probe setup (FD-TPS) [9]-[13], the time-domain two-probe setup (TD-TPS) [14]-[17] and the frequency-domain single-probe setup (FD-SPS) [18]-[20]. Besides, the multi-probe setup (MPS) was also designed but it is for simultaneous measurement of in-circuit impedances of multiple SUTs in multi-branches powered by the same power source [21], [22]. Although the principle of each setup has been introduced, a comprehensive survey and comparison of these setups have not been found in the literature. To apply these setups effectively, this paper reviews the FD-TPS, TD-TPS and FD-SPS in terms of their characteristics, limitations, and applications. In addition, recommendations for future research are also presented.

## II. Frequency-Domain Two-Probe Setup

The FD-TPS was first reported for power line in-circuit impedance measurement [9]. It was later improved and extended to many other EMC applications, such as electromagnetic interference (EMI) filter design of power converters [10] and radiated emission estimation from photovoltaic (PV) systems [11]. As shown in Fig. 1(a), the FD-TPS consists of two clamp-on inductive probes (IP1 and IP2) and a vector network analyzer (VNA). To measure the in-circuit impedance of the SUT ($Z_X$) energized by a power source, IP1 and IP2 are clamped on the wiring connection of the SUT with the position denoted as *c-c′*. One port of the VNA injects a stepped swept-sine excitation signal into the SUT through an inductive probe while the other port measures the response through another inductive probe. Fig. 1(b) shows its cascaded two-port network (C2PN) equivalent circuit [12]. $N_{IP1}$ and $N_{IP2}$ represent the two-port networks of the IP1 and IP2 with the respective wiring being clamped. $N_X$ is the two-port network of the SUT. $N$ is the resultant two-port network of $N_{IP1}$, $N_X$ and $N_{IP2}$. Expressing these two-port networks in terms of transmission (ABCD) parameters, we obtain:

$$\begin{bmatrix} A & B \\ C & D \end{bmatrix} = \begin{bmatrix} A_{IP1} & B_{IP1} \\ C_{IP1} & D_{IP1} \end{bmatrix} \begin{bmatrix} A_X & B_X \\ C_X & D_X \end{bmatrix} \begin{bmatrix} A_{IP2} & B_{IP2} \\ C_{IP2} & D_{IP2} \end{bmatrix} \quad (1)$$

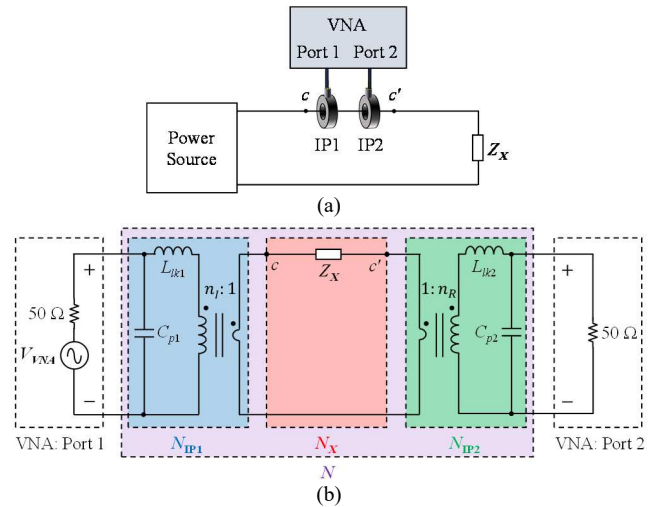

Fig. 1. (a) FD-TPS. (b). C2PN equivalent circuit.

From (1), once the ABCD parameters of $N$, $N_{IP1}$ and $N_{IP2}$ are known, the ABCD parameters of $N_X$ can be solved. Since $B_X = Z_X$ [23], $Z_X$ can be obtained finally. The ABCD parameters of $N$ can be derived from the measured scattering (S) parameters using the VNA. To extract the ABCD parameters of $N_{IP1}$ and $N_{IP2}$, a specific test fixture and a characterization process have been proposed [12].

## III. Time-Domain Two-Probe Setup

The FD-TPS is usually used for time-invariant in-circuit impedance measurement. To perform time-variant in-circuit impedance measurement, the TD-TPS was developed. The TD-TPS was first reported for time-variant in-circuit impedance monitoring of switching circuits [14]. It was later employed for voltage-dependent capacitances extraction of power semiconductor devices [8] and stator insulation faults detection of induction motors [17]. As shown in Fig. 2(a), the TD-TPS consists of one injecting inductive probe (IIP), one receiving inductive probe (RIP), and a computer-controlled signal generation and acquisition system (SGAS). A single-sine excitation signal produced by the signal generation card (SGC) is injected into the SUT through the IIP, and its response is monitored through the RIP. The two channels (CH1 and CH2) of the signal acquisition card (SAC) measure the excitation signal voltage at the IIP ($V_1$) and the response signal voltage at the RIP ($V_2$), respectively. Fig. 2(b) shows its three-port network (3PN) equivalent circuit [15]. Compared with C2PN equivalent circuit, the 3PN equivalent circuit considers the influence of probe-to-probe coupling between the two inductive probes. Based on 3PN equivalent circuit, $Z_X$ can be expressed in terms of $V_1$ and $V_2$ as follows:

$$Z_X = \frac{a_3 \cdot (V_1/V_2) - a_2}{-V_1/V_2 + a_1} \quad (2)$$

where $a_1$, $a_2$ and $a_3$ are the frequency-dependent intrinsic parameters of TD-TPS, and their full expressions are:

$$a_1 = \frac{Z_{11}}{Z_{21}} \cdot \left(1 - \frac{Z_{22}}{Z_{CH2}}\right) + \frac{Z_{12}}{Z_{CH2}} \quad (3)$$

$$a_2 = \left(Z_{13} - \frac{Z_{11}Z_{23}}{Z_{21}}\right) \cdot \left[\frac{Z_{31}}{Z_{21}} \cdot \left(1 - \frac{Z_{22}}{Z_{CH2}}\right) + \frac{Z_{32}}{Z_{CH2}}\right] - \left(Z_{33} - \frac{Z_{31}Z_{23}}{Z_{21}}\right) \cdot \left[\frac{Z_{11}}{Z_{21}} \cdot \left(1 - \frac{Z_{22}}{Z_{CH2}}\right) + \frac{Z_{12}}{Z_{CH2}}\right] \quad (4)$$

$$a_3 = \frac{Z_{31} \cdot Z_{23}}{Z_{21}} - Z_{33} \quad (5)$$

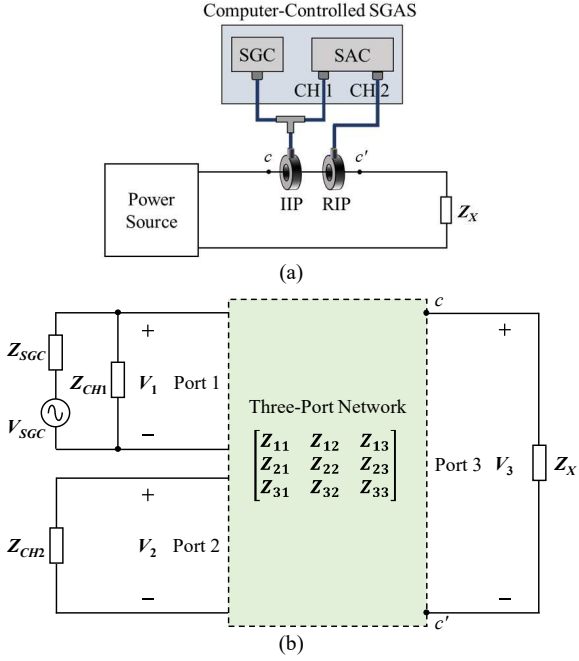

Fig. 2. (a) TD-TPS. (b) 3PN equivalent circuit.

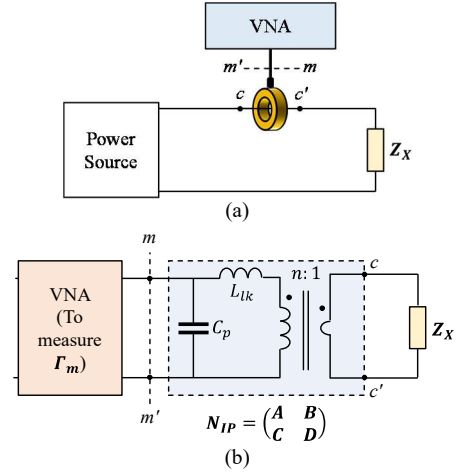

Fig. 3. (a) FD-SPS. (b). 2PN equivalent circuit.

In (3)-(5), $Z_{ij}$ ($i, j = 1, 2, 3$) are impedance parameters of the three-port network. $Z_{CH1}$ and $Z_{CH2}$ are the internal impedances of CH1 and CH2, respectively. By executing a pre-measurement calibration procedure [16], $a_1$, $a_2$ and $a_3$ of a given TD-TPS can be determined so that $Z_X$ can be derived from (2) based on the measured $V_1$ and $V_2$. To continuously measure $V_1$ and $V_2$, their time-domain counterparts (i.e., $v_1$ and $v_2$) are sampled and a moving window discrete Fourier transform (MWDFT) algorithm is applied [14]. Based on the continuously measured $V_1$ and $V_2$, the time-variant $Z_X$ can be extracted.

For the applications where strong electrical noise and power surges are present, both FD-TPS and TD-TPS can incorporate a signal amplification and protection (SAP) module to improve its signal-to-noise ratio (SNR) and enhance its ruggedness [13], [17].

IV. FREQUENCY-DOMAIN SINGLE-PROBE SETUP

The FD-SPS was developed to eliminate the probe-to-probe coupling issue encountered by the FD-TPS and TD-TPS [18]. The FD-SPS has been employed to extract the common-mode (CM) and differential-mode (DM) impedances of motor drive systems [19], [20]. As shown in Fig. 3(a), it consists of one clamp-on inductive probe and a frequency-domain measurement instrument (e.g. VNA). It can also incorporate a SAP module for applications where strong noise and power surges are present [18]. To measure $Z_X$, the VNA generates a stepped swept-sine excitation through its internal signal source and provides the reflection coefficient at $m$-$m'$ denoted by $\Gamma_m$. Fig. 3(b) shows the two-port network (2PN) equivalent circuit. $N_{IP}$ represents the two-port network of the inducive probe with the wire being clamped. Expressing $N_{IP}$ using ABCD parameters, the relationship between $Z_X$ and $\Gamma_m$ can be established by:

$$Z_X = \frac{k_1 \cdot \Gamma_m + k_2}{\Gamma_m + k_3} \quad (6)$$

where

$$k_1 = -\frac{Z_0 \cdot D + B}{Z_0 \cdot C + A} \quad (7)$$

TABLE I
SUMMARY OF MEASUREMENT SETUPS OF INDUCTIVE COUPLING APPROACH

| Measurement Setups | Excitation Signal | Measured Parameter | Time-Variant Impedance Measurement | Probe-to-Probe Coupling |
|---|---|---|---|---|
| FD-TPS | Stepped swept-sine | S-parameter | Inappropriate | Have |
| TD-TPS | Single-sine | Time-domain voltage | Appropriate | Don't have |
| FD-SPS | Stepped swept-sine | S-parameter | Inappropriate | Have |

$$k_2 = -\frac{Z_0 \cdot D - B}{Z_0 \cdot C + A} \quad (8)$$

$$k_3 = \frac{Z_0 \cdot C - A}{Z_0 \cdot C + A} \quad (9)$$

$Z_0$ represents the VNA's reference impedance. $k_1$, $k_2$ and $k_3$ are frequency-dependent intrinsic parameters of the SPS. By performing a calibration prior to the in-circuit measurement [18], $k_1$, $k_2$ and $k_3$ of a given SPS can be determined so that $Z_X$ can be obtained from (6) based on the measured $\Gamma_m$. It should be mentioned that the FD-SPS is usually applied for time-invariant in-circuit impedance measurement.

V. SUMMARY AND CONCLUSION

Table I summarizes the above-mentioned measurement setups from the following four aspects: the excitation signal, measured parameter, time-variant in-circuit impedance measurement, and probe-to-probe coupling. From the table, all three setups perform the measurement via stepped swept-sine or single-sine excitation, where the measurement takes place at only a single frequency at a time. Future research will explore the multi-sine excitation for multi-frequency simultaneous measurement of in-circuit impedance.

REFERENCES

[1] H. Chen, Y. Yan, and H. Zhao, "Extraction of common-mode impedance of an inverter-fed induction motor," *IEEE Trans. Electromagn. Compat.,* vol. 58, no. 2, pp. 599 - 606, Apr. 2016.

[2] Y. Liu, S. Jiang, H. Wang, G. Wang, J. Yin, and J. Peng, "EMI filter design of single-phase SiC MOSFET inverter with extracted noise source impedance," *IEEE Electromagn. Compat. Mag.,* vol. 8, no. 1, pp. 45–53, Jan.-Mar. 2019.

[3] J. Yao, S. Wang, and H. Zhao, "Measurement techniques of common mode currents, voltages, and impedances in a flyback converter for radiated EMI diagnosis," *IEEE Trans. Electromagn. Compat.,* vol. 61, no. 6, pp. 1997-2005, Dec. 2019.

[4] E. Mazzola, F. Grassi, and A. Amaducci, "Novel measurement procedure for switched-mode power supply modal impedances," *IEEE Trans. Electromagn. Compat.,* vol. 62, no. 4, pp. 1349-1357, Aug. 2020.

[5] L. Wan *et al.*, "Assessment of validity conditions for black-box EMI modelling of DC/DC converters," in *Proc. IEEE Int. Joint EMC/SI/PI, EMC Europe Symp.*, Raleigh, NC, USA, 2021, pp. 581-585.

[6] H. Hu, P. Pan, Y. Song, and Z. He, "A novel controlled frequency band impedance measurement approach for single-phase railway traction power system," *IEEE Trans. Ind. Electron.,* vol. 67, no. 1, pp. 244-253, Jan. 2020.

[7] X. Shang, D. Su, H. Xu, and Z. Peng, "A noise source impedance extraction method for operating SMPS using modified LISN and simplified calibration procedure," *IEEE Trans. Power Electron.,* vol. 32, no. 6, pp. 4132-4139, Jun. 2017.

[8] Z. Zhao *et al.*, "Voltage-dependent capacitance extraction of SiC power MOSFETs using inductively coupled in-circuit impedance measurement technique," *IEEE Trans. Electromagn. Compat.,* vol. 61, no. 4, pp. 1322-1328, Aug. 2019.

[9] R. A. Southwick and W. C. Dolle, "Line impedance measuring instrumentation utilizing current probe coupling," *IEEE Trans. Electromagn. Compat.,* vol. EMC-13, no. 4, pp. 31-36, Nov. 1971.

[10] V. Tarateeraseth, K. Y. See, F. G. Canavero, and R. W.-Y. Chang, "Systematic electromagnetic interference filter design based on information from in-circuit impedance measurements," *IEEE Trans. Electromagn. Compat.,* vol. 52, no. 3, pp. 588-598, Aug. 2010.

[11] M. Prajapati, F. Fan, Z. Zhao, and K. Y. See, "Estimation of radiated emissions from PV system through black box approach," *IEEE Trans. Instrum. Meas.,* vol. 70, 2021, Art no. 9004304, doi: 10.1109/TIM.2021.3102752.

[12] K. Li *et al.*, "Inductive coupled in-circuit impedance monitoring of electrical system using two-port ABCD network approach," *IEEE Trans. Instrum. Meas.,* vol. 64, no. 9, pp. 2489-2495, Sep. 2015.

[13] F. Fan, K. Y. See, X. Liu, K. Li, and A. K. Gupta, "Systematic common-mode filter design for inverter-driven motor system based on in-circuit impedance extraction," *IEEE Trans. Electromagn. Compat.,* vol. 62, no. 5, pp. 1711 - 1722, Oct. 2020.

[14] Z. Zhao *et al.*, "Time-variant in-circuit impedance monitoring based on the inductive coupling method," *IEEE Trans. Instrum. Meas.,* vol. 68, no. 1, pp. 169-176, Jan. 2019.

[15] Z. Zhao *et al.*, "Eliminating the effect of probe-to-probe coupling in inductive coupling method for in-circuit impedance measurement," *IEEE Trans. Instrum. Meas.,* vol. 70, 2021, Art no. 1000908, doi: 10.1109/TIM.2020.3013688.

[16] Z. Zhao, A. Weerasinghe, Q. Sun, F. Fan, and K. Y. See, "Improved calibration technique for two-probe setup to enhance its in-circuit impedance measurement accuracy," *Measurement,* vol. 185, 2021, Art no. 110007, doi: 10.1016/j.measurement.2021.110007.

[17] Z. Zhao, F. Fan, W. Wang, Y. Liu, and K. Y. See, "Detection of stator interturn short-circuit faults in inverter-fed induction motors by online common-mode impedance monitoring," *IEEE Trans. Instrum. Meas.,* vol. 70, 2021, Art no. 3513110, doi: 10.1109/TIM.2021.3066193.

[18] A. Weerasinghe *et al.*, "Single-probe inductively coupled in-circuit impedance measurement," *IEEE Trans. Electromagn. Compat.,* vol. 64, no. 1, pp. 2-10, Feb. 2022.

[19] Z. Zhao, F. Fan, A. Weerasinghe, P. Tu, and K. Y. See, "Measurement of in-circuit common-mode impedance at the ac input of a motor drive system," in *Proc. IEEE Asia-Pacific Symp. Electromagn. Compat. (APEMC)*, Bali, Indonesia, 2021, pp. 1-4.

[20] A. Weerasinghe, Z. Zhao, F. Fan, P. Tu, and K. Y. See, "In-circuit differential-mode impedance extraction at the AC input of a motor drive system," in *Proc. IEEE Asia-Pacific Symp. Electromagn. Compat. (APEMC)*, Bali, Indonesia, 2021, pp. 1-4.

[21] K. Li, A. Videt, and N. Idir, "Multiprobe measurement method for voltage-dependent capacitances of power semiconductor devices in high voltage," *IEEE Trans. Power Electron.,* vol. 28, no. 11, pp. 5414-5422, Nov. 2013.

[22] Z. Zhao and K. Y. See, "A multiprobe inductive coupling method for online impedance measurement of electrical devices distributed in multibranch cables," *IEEE Trans. Instrum. Meas.,* vol. 69, no. 9, pp. 5975-5977, Sep. 2020.

[23] D. M. Pozar, *Microwave engineering*. Hoboken, NJ, USA: Wiley, 2012, pp. 188-194.